\documentclass[prd,showpacs,showkeys,nofootinbib,eqsecnum,floatfix,fleqn,
               preprint,12pt,tightenlines]{revtex4} %for preprint

\usepackage{amsmath,amssymb,revsymb,graphicx,dcolumn}

\newcommand{\beq}{\begin{equation}}
\newcommand{\eeq}{\end{equation}}
\newcommand{\beqa}{\begin{eqnarray}}
\newcommand{\eeqa}{\end{eqnarray}}
\newcommand{\bsubeqs}{\begin{subequations}}
\newcommand{\esubeqs}{\end{subequations}}

\newcommand{\half}{{\textstyle \frac{1}{2}}}    % 1/2

\newcommand{\lambdaOneTwo}{\lambda_{12}}  %% zeta=lambda10 in MATH
\newcommand{\lambdaTwoOne}{\lambda_{21}}

\begin{document}

%\preprint{arXiv:1101.1281\;(\version)}
\noindent Phys. Rev. D 84, 023011 (2011) \hfill arXiv:1101.1281\newline
\vspace*{2mm}
\title{Effective cosmological constant from TeV--scale physics:\\
Simple field-theoretic model\vspace*{5mm}}
\author{F.R. Klinkhamer}
\email{frans.klinkhamer@kit.edu}
\affiliation{\mbox{Institute for Theoretical Physics, University of Karlsruhe,}
Karlsruhe Institute of Technology, 76128 Karlsruhe, Germany\\}

\begin{abstract}
%%\vspace*{2.5mm} %%out for twocolumn
\noindent
Adopting the $q$--theory approach to the cosmological constant
problem, a simple field-theoretic model is presented
which generates an effective cosmological constant
(remnant vacuum energy density) of the observed order of magnitude,
$\Lambda_\text{eff} \sim (\text{meV})^4$,
if there exist new $\text{TeV}$--scale ultramassive particles
with electroweak interactions.
The model is simple, in the sense that it involves only a few types   %%%%\marginpar{ZZZ}
of fields and two energy scales, the gravitational energy
scale $E_\text{Planck}\sim 10^{15}\;\text{TeV}$ and the
electroweak (new-physics) energy scale $E_\text{ew}\sim 1-10\;\text{TeV}$.
\end{abstract}

  %%FRK \hfill (June 13, 2011;\,\version)
\maketitle

\section{Introduction}
\label{sec:Introduction}

The main cosmological constant problem (CCP1) can be phrased as
follows (see, e.g., the review~\cite{Weinberg1988}):
\emph{why do the quantum fields in the vacuum  not produce
naturally a large absolute value for the cosmological constant
$\mathit{\Lambda}$ in the Einstein field equations or, practically,
why is the measured value of $|\mathit{\Lambda}|^{1/4}$ very
much smaller than the known energy scales of elementary
particle physics?}
One possible solution relies on the so-called $q$--theory
approach~\cite{KV2008-statics,KV2008-dynamics,KV2009-CCP1},
which provides a compensation-type solution of CCP1
by self-adjustment of the $q(x)$ field.
This $q$ field, considered to describe ultrahigh-energy
microscopic degrees of freedom, must be of a very special type,
being relativistic and conserved in the equilibrium state
(Minkowski spacetime). In fact, the equilibrium value $q_{0}$
is constant over spacetime. This property allows for
the study of the macroscopic equations in terms of $q_{0}$,  
which, in particular, give a vanishing
gravitating vacuum energy density
(provided there is no external pressure):
$\rho_{V}(q_{0}) \equiv \Lambda_{0} =0$.

In this way, the zero-point energies
of the standard-model fields in the equilibrium vacuum state
can be compensated exactly by contributions from the
microscopic degrees of freedom at a higher energy scale,
without need to know the detailed microscopic theory. Having
provided a possible explanation of the vanishing
cosmological constant $\Lambda_{0}$ in the equilibrium state,
the next task is to explain the measured small but nonzero value
of the effective cosmological constant $\Lambda_\text{eff}$
in the present expanding (nonequilibrium) Universe.
The search for the explanation of this last number,
$\Lambda_\text{eff}$, has been
called the second cosmological constant problem (CCP2),
even though the term `puzzle' is perhaps more appropriate.

In the early phase of the history of the Universe
(close to the Planck epoch), the $q$--theory dynamical
equations~\cite{KV2008-dynamics,KV2009-CCP1} show that
the gravitating vacuum energy density $\rho_{V}[q(t)]$
is rapidly relaxed to zero.
What happens next depends on the details
of the particle-physics theory, in particular, the theory
at the TeV energy scale~\cite{ArkaniHamed-etal2000}.
If there exist new ultramassive TeV--scale particles with electroweak
interactions, the presence of these ultramassive particles
affects the expansion rate of the Universe, and this change of the
expansion rate ``kicks'' $\rho_{V}(t)$ away from
zero~\cite{KV2009-electroweak,K2010-Lambda-TeV}.
The maximal value of $\rho_{V}(t)$ is of the order
of $(\text{meV})^4$, consistent with the value suggested by
observational cosmology. The problem, however, is that $\rho_{V}(t)$
can be expected to drop to zero again if the ultramassive particles
ultimately disappear.

Possible quantum-dissipative effects~\cite{KV2009-electroweak}
may lead to a freezing of the gravitating vacuum energy
density $\rho_{V}$, but this does not guarantee
an asymptotic approach to a standard de-Sitter universe
(consistent with the $\Lambda$CDM model of the present Universe).
In fact, there appears to be a
potential mismatch~\cite{K2010-Lambda-TeV}
between the $q$--theory dynamical equations and those of standard
general relativity with a nonzero cosmological constant.
It is possible to modify the relevant $q$--theory dynamical equation
by hand~\cite{K2010-Lambda-TeV}, but such a procedure
is unsatisfactory.

In this article, it is shown that it is possible to remain
entirely within the framework of $q$--theory
by allowing for a nontrivial interaction between, on the one hand,
the $q$--field and, on the other hand,
the matter and gravitational fields.
The model is remarkably simple and has one crucial ingredient,
which, ultimately, needs to be derived from the underlying microscopic
theory (assuming that the model is relevant).
The purpose of this article is to present the simple model
and to perform a numerical calculation in order to make sure that
there is indeed a nonzero remnant vacuum energy density,
leaving an extensive discussion of the systematics
to a future publication.

In order to place the present work in context,
the reader is referred to a recent review article
on $q$--theory and the evolution of vacuum energy density in
cosmology~\cite{KV2011-review}.
At this moment, it may also be helpful to comment on
how the results of the present article compare with those
of so-called quintessence models from a dynamic scalar
field~\cite{PeeblesRatra1988,Zlatev-etal1998}
(see, e.g., Sec.~8 of the review~\cite{SahniStarobinsky2000}
for further discussion and references).
There are, at least, two basic differences.

First, the standard quintessence models have a
fundamental canonical scalar field $\phi(x)$ and no natural
explanation of CCP1 (see, e.g., Sec.~VI of Ref.~\cite{Weinberg1988}),
whereas the \mbox{(pseudo-)scalar} $q(x)$ field is nonfundamental
and special, in order to provide a possible
solution of CCP1 as explained in Ref.~\cite{KV2009-CCP1}.
(Remark that quintessence models may still provided valuable
insights into CCP2, if CCP1 can be assumed to be solved.)
Second, the ``dark energy''
from dynamic scalar fields $\phi$ has generally
an equation-of-state (EOS) parameter $w \equiv P/\rho \ne -1$,
whereas the vacuum energy density of $q$--theory in its simplest form
has $w=-1$ exactly. As regards $q$--theory, both points
are clarified by considering explicit realizations,
see Refs.~\cite{KV2008-statics,KV2008-dynamics,KV2009-CCP1}
and Sec.~\ref{subsec:General-properties}.

In short, the ultimate goal is to find an explanation of \emph{both}
CCP1 and CCP2. This article tries to make a modest contribution
towards reaching that goal.

%%\newpage%%tmp
\section{Field-theoretic model}
\label{sec:Field-theoretic-model}

\subsection{General properties}
\label{subsec:General-properties}

Consider two real scalars: an ultramassive scalar field $\sigma$
with mass $M=E_\text{ew}\sim \text{TeV}$
and a strictly massless scalar field $\psi$.
Let $\sigma$ now exists in $N_{1}$ identical copies
and $\psi$ in $N_{2}$ copies. Then, the
scalars $\psi_{c}$, for $c=1,\,\ldots,\,N_{2}$
and $N_{2} \sim 10^2$, may correspond to the particles of the
Standard Model\footnote{\label{ftn:SMlike}In principle,
it is no problem to extend
the theory of $\psi_{c}$ scalars to the one of the standard model
with gauge and Higgs bosonic fields
and quark and lepton fermionic fields.
Actually, the fermions give the largest contribution to the
total number of degrees of freedom in the standard model,
hence the suggestive notation $\psi_{c}$
for the corresponding scalars of the simplified model.}
and the scalars $\sigma_{b}$, for $b=1,\,\ldots,\,N_{1}$,
to the particles of new TeV-scale physics
(perhaps with $N_1=N_2 \sim 10^2$ resulting from broken
supersymmetry~\cite{WessZumino1974,FayetFerrara1976}).
From now on, the indices $b,c$ will be kept implicit
by using an inner-product notation with
$\sigma\cdot\sigma \equiv \sum_{b}\,\sigma_{b}\,\sigma_{b}$
and $\psi\cdot\psi \equiv \sum_{c}\,\psi_{c}\,\psi_{c}$.

This article employs the framework of $q$--theory,
possibly viewed as an effective theory in\-cor\-po\-ra\-ting
quantum effects of vacuum-matter interactions.
Following Refs.~\cite{KV2008-dynamics,KV2009-CCP1},
the $q$--theory is considered to be realized via a three-form
gauge field $A$~\cite{DuffNieuwenhuizen1980,Aurilia-etal1980}.
The macroscopic effective action of the relevant ``ultraviolet''
fields (here, $A$) and the ``infrared'' fields
(here, $g$, $\sigma$, and $\psi$)
is taken to be of the following form:
\bsubeqs\label{eq:Seff-q2def-Fdef}
\beqa
%\hspace*{-10mm} %%twocolumn
S_{\text{eff},\,T}[A,\, g,\,\sigma,\,\psi]&=&
\int_{\mathbb{R}^4} \,d^4x\, \sqrt{-\text{det}(g)}\,
\Big( K_{T}(q)\,R[g]
%\nonumber\\\hspace*{-0mm}&&  %%twocolumn
+ \epsilon_{V}(q)+\mathcal{L}^{(M)}_{\text{eff},\,T}[\sigma,\psi,g] \Big)\,,
\label{eq:Seff}
\\[2mm]
%\hspace*{-10mm} %%twocolumn
q &\equiv&
-\frac{1}{24}\; \epsilon^{\alpha\beta\gamma\delta}\,
\nabla_{\alpha}A_{\beta\gamma\delta}\,/\,\sqrt{-\text{det}(g)}\,,
\label{eq:qdef}
\eeqa
\esubeqs
where $R[g]$ is the Ricci curvature scalar obtained from the metric
$g_{\alpha\beta}$,
$\epsilon_{\alpha\beta\gamma\delta}$ the Levi-Civita tensor density,
and $\nabla_\alpha$  the standard covariant derivative.
The effective gravitational-coupling parameter $K_{T}$ in
\eqref{eq:Seff}
is an even function of the vacuum variable $q=q[A,\, g]$ and
an indirect function of the matter fields manifesting itself
as a dependence on the temperature $T$ (see below).
The energy density $\epsilon_{V}(q)$
is assumed to be a generic even function of the vacuum variable $q$,
that is, a function different
from the quadratic $\half\,q^2$ corresponding to a
Maxwell-type theory~\cite{DuffNieuwenhuizen1980,Aurilia-etal1980}.
Here, and in the following, set
$\hbar=c=k=1$ and take the metric signature $(-1,\, 1,\, 1,\, 1)$.

The $q$--dependence of $K$, first introduced in Ref.~\cite{KV2008-dynamics},
allows for a continuous time dependence of $q$ in a cosmological context.
This behavior differs from the stepwise evolution of
$q$ in the Brown--Teitelboim mechanism which operates via
quantum tunneling~\cite{BrownTeitelboim1987,BoussoPolchinski2000}.
The physical motivation for having $K(q)$ is that $q$ is considered
to be one of the variables which characterize the quantum vacuum
and that, therefore, $q$ can be expected to determine all
``constants'' of the low-energy theory, one of which controls the
gravitational coupling. The possible $q$--dependence of the
``constants'' in the matter Lagrange density $\mathcal{L}^{(M)}$
of \eqref{eq:Seff} is neglected for simplicity.
As it stands, the action \eqref{eq:Seff} corresponds
to a type of Brans--Dicke model~\cite{BransDicke1961}
with a special nonfundamental pseudoscalar field $q$.

The nonfundamental pseudoscalar field $q(x)$ from
\eqref{eq:qdef} has been called the
``quinta-essentia'' field in Ref.~\cite{KV2009-CCP1},
in order to distinguish it from the
fundamental scalar field $\phi(x)$ of typical quintessence
models~\cite{PeeblesRatra1988,Zlatev-etal1998,SahniStarobinsky2000}.
The explicit realization of $q$ via \eqref{eq:Seff-q2def-Fdef}
clarifies the two points mentioned in the penultimate paragraph
of Sec.~\ref{sec:Introduction}, $q$ being nonfundamental and
giving rise to a dynamical vacuum energy density with
EOS parameter $w_V=-1$.
Note that, for this particular realization of $q$--theory,
the mass dimension of $q$ equals 2.
For different realizations, $q$ may have different mass dimensions
and intrinsic parities.

\subsection{Specific properties}
\label{subsec:Specific-properties}

As explained in the Introduction,
the main focus is on the freezing mechanism of the
vacuum energy density and the scalar Lagrange density is kept
as simple as possible, only asking that it produces
massive $\sigma$ scalars at an appropriate epoch in the
history of the Universe (temperatures of order $E_\text{ew}$).
Specifically, take the following Lagrange density with a single
quartic coupling term between the two types of scalars:
\bsubeqs\label{eq:Lscalar-M-rhoV-KAnsatz}
\beqa
\hspace*{-2mm}
\mathcal{L}^{(M)}_{\text{eff},\,T}[\sigma,\psi,g]
&=&
 \frac{1}{2}\, \partial_\alpha\psi\,\cdot\, \partial^\alpha\psi
%\nonumber\\&&  %%twocolumn
+\frac{1}{2}\, \partial_\alpha\sigma\,\cdot\, \partial^\alpha\sigma
+\frac{1}{2}\,M^2\, (\sigma\cdot\sigma)
%\nonumber\\&&  %%twocolumn
+g_{T}\,(\psi\cdot\psi)\,(\sigma\cdot\sigma)\,,
\label{eq:Lscalar}\\[2mm]
\hspace*{-2mm}
M&=& E_\text{ew} \,,
\label{eq:M-Ansatz}\\[2mm]
\hspace*{-2mm}
g_{T}&=&
\left\{\begin{array}{l}
0\hspace*{30.5mm} \quad\,\text{for}\; T > T_{c,\,g}  \,,\\[1mm]
g_0\; \Big(1-(T/T_{c,\,g})^2\Big)\quad\text{for}\; T \leq T_{c,\,g} \,,
\end{array}\right.
%\nonumber\\&&  %%twocolumn
\label{eq:gT-Ansatz}
\eeqa
where $T_{c,\,g}$ is a critical temperature of order $E_\text{ew}$,
above which the scalar interactions are suppressed.
The nontrivial temperature behavior of \eqref{eq:gT-Ansatz} may
effectively result from an interaction term
$(\psi\cdot\psi)\,(\sigma\cdot\sigma)\,
 \widetilde{\chi}^{\,2}/(E_\text{ew})^2$
in an extended theory where a single
$\widetilde{\chi}$ scalar picks up an expectation value
at temperatures below a second-order continuous phase transition
(see, e.g., Sec.~4.4 in Ref.~\cite{Mukhanov2005}).
But, most likely, this phase-transition explanation of the
postulated behavior \eqref{eq:gT-Ansatz}
does not need to be taken literally:
$g_{T}$ may very well have an entirely different origin,
provided the model is relevant
at all.\footnote{\label{ftn:rad-corr}The quartic interaction
term in \eqref{eq:Lscalar} leads to radiative corrections for the
low-energy theory of the scalars $\psi_{c}$ (which mimic the
standard model particles as mentioned in Ftn.~\ref{ftn:SMlike}),
but these corrections are suppressed by the large masses of the
$\sigma_{b}$ scalars (in the simple model, all masses are taken
to be equal to $M$).
Still, radiative corrections may provide a valuable alternative
to direct searches if all masses of the new $\sigma$--type
particles are of the order of several $\text{TeV}$ or more.}

The gravitating vacuum energy density near equilibrium ($q=q_0$) is
taken to be quadratic~\cite{KV2009-electroweak,K2010-Lambda-TeV}
\beqa
\rho_{V}(q) &\equiv&
\epsilon(q)-\mu_0\, q =\frac{1}{2}\,\big( q-q_0 \big)^2\,,
\label{eq:rhoV-Ansatz}
\eeqa
with $\epsilon=\epsilon_{V}$ for the scalar theory
\eqref{eq:Lscalar} as it stands and an appropriate constant
value $q_0$ of the dynamic $q$--field (or an appropriate
value $\mu_0$ of the corresponding
chemical potential $\mu$;
see Refs.~\cite{KV2008-statics,KV2008-dynamics,KV2009-CCP1,KV2011-review}
for further discussion).

The really new input for model \eqref{eq:Seff} is the following
\textit{Ansatz} for the effective gravitational-coupling parameter:
\beqa
K_{T}(q)
&=&
\frac{1}{2}\, q_0 + \frac{1}{2}\,\big(q-q_0\big)\;
\theta\Big[ T/T_{c,\,K}^{\,(+)} -1  \Big]
=             %%preprint
%\nonumber\\&=&  %%twocolumn
\left\{\begin{array}{l}
  q/2_{\phantom{0}}   \;\;\;\; \text{for}\;\;\;\;
  T >    T_{c,\,K}^{\,(+)}  \,, \\[2mm]
  q_0/2 \;\;\;\; \text{for}\;\;\;\;
  T \leq T_{c,\,K}^{\,(+)} \,,
\end{array}\right.
\label{eq:KAnsatz}
\eeqa
with the step function
\beqa
\theta[x]&=&
\left\{\begin{array}{l}
  1   \;\;\;\; \text{for}\;\;\;\;  x >    0  \,, \\[0mm]
  0   \;\;\;\; \text{for}\;\;\;\;  x \leq 0 \,.
\end{array}\right.
\label{eq:theta-def}
\eeqa
For the model universe to be discussed in Sec.~\ref{sec:Cosmology}
having a temperature decreasing with cosmic time,
$K_{T}$ is a nontrivial function of $q$ above a critical temperature $T_{c,\,K}^{\,(+)}$
and a constant below $T_{c,\,K}^{\,(+)}$.
A possible physical realization of the
corresponding ``first-order phase transition'' will be given
in Sec.~\ref{subsec:Additional-remarks}.
Here, two brief remarks suffice. First, there may be
hysteresis-type effects, and the suffix `$(+)$' on the critical
temperature is to indicate that the transition is approached
from the high-temperature side.
Second, there are the following
assumptions on the critical temperatures entering
\eqref{eq:gT-Ansatz} and \eqref{eq:KAnsatz}:
\beq
T_{c,\,g}         = \text{O}(E_\text{ew})\,,\quad
T_{c,\,K}^{\,(+)}  =  \text{O}(E_\text{ew})\,,\quad
T_{c,\,g} > T_{c,\,K}^{\,(+)}\,.
\label{eq:Tc-assumptions}
\eeq
\esubeqs

The constant $q_0$ in \eqref{eq:KAnsatz}, relevant
at zero temperature, is proportional to
the inverse of Newton's gravitational constant, specifically
\bsubeqs\label{eq:q0-def-xi-def}
\beq\label{eq:q0def-EPlanck}
q_0=1/(8\pi\,G_N) \equiv (E_\text{Planck})^2
\approx (2.44\times 10^{18}\:\text{GeV})^2\,.
\eeq
Given the energy scale $E_\text{ew}$ from \eqref{eq:M-Ansatz}
and \eqref{eq:Tc-assumptions},
a single dimensionless parameter characterizes the
theory, namely, the ratio of the two energy-density scales,
\beq\label{eq:xi-def}
\xi\equiv  (E_\text{Planck}/E_\text{ew})^4\,,
\eeq
\esubeqs
which is approximately $10^{60}$
for $E_\text{ew}\approx 2.44\:\text{TeV}$.

\subsection{Additional remarks}
\label{subsec:Additional-remarks}

The previous subsection has given the detailed description
of the field-theoretic model \eqref{eq:Seff}. Still outstanding
is the promised physical realization
of the gravitational-coupling \textit{Ansatz} \eqref{eq:KAnsatz}.
The particular realization relies on the possibility of having
symmetry restoration at low temperatures and symmetry breaking
at high temperatures, the opposite of what is the case
in most systems. This possibility has been discussed in, e.g.,
Ref.~\cite{Weinberg1974} and it is the easiest to just follow
Example 3 of Sec. IV of that article.

The argument, then,
proceeds in four steps. First, start from a scalar theory
with global $O(n)\times O(n)$ symmetry~\cite{Weinberg1974},
where the scalar fields
are denoted as $\chi_A$ and $\eta_a$, respectively, with both indices
$A$ and $a$ running over $1,\,\ldots,\,n$
[again, an inner-product notation will be used,
$\chi\cdot\chi    \equiv \sum_{A}\,\chi_A\,\chi_A$
and $\eta\cdot\eta\equiv \sum_{a}\,\eta_a\,\eta_a$].
Second, take the parameters in the zero-temperature
potential in such a way as to give the following
symmetry-breaking pattern in a finite-temperature context:
\bsubeqs\label{eq:K-argument-SSB-KRterm}
\beq\label{eq:K-argument-SSB}
O(n)\times O(n)\;\Big|_{T=0}
\;\;\stackrel{1\text{PT}}{\longrightarrow}\;\;
O(n)\times O(n-1)\;\Big|_{T=\infty}\;,
\eeq
which implies that the $\eta$ scalars develop an expectation value
at high temperatures. As indicated by the superscript on the
arrow in \eqref{eq:K-argument-SSB}, the finite-temperature
phase transition is arranged to be
first-order.\footnote{\label{ftn:Vchieta}Specifically,
the potential $P(\chi,\,\eta)$
is given on p.~3367, left column of Ref.~\cite{Weinberg1974}
with mass-square parameters now positive and of order $(E_\text{ew})^2$,
to which are added the following ``cubic'' terms:
$g_{1}\,E_\text{ew}\,(\chi\cdot\chi)^{3/2}$
and $g_{2}\,E_\text{ew}\,(\eta\cdot\eta)^{3/2}$.
The coupling constants are taken as in
the fourth unnumbered equation on p.~3368, left column of the
same reference, together with, for example, $g_{1}<0$ and
$g_{2}<0$. These cubic terms and the finite-temperature corrections
of order $T\,\eta^3$ can give a
first-order (discontinuous) phase transition~\cite{Mukhanov2005}.
The first-order nature of the
phase transition \eqref{eq:K-argument-SSB} may also have
other origins, see Endnote~[24] of Ref.~\cite{Weinberg1974}.}
Third, consider the following hypothetical interaction term
in the effective Lagrange density:
\beq\label{eq:K-argument-KRterm}
K_{0}\:R +  \big(q/2-K_{0}\big)\;
\frac{\chi\cdot\chi+\eta\cdot\eta}
     {(E_\text{ew})^2+\chi\cdot\chi+\eta\cdot\eta}\;R\,,
\eeq
\esubeqs
where the fraction is close to 1 for field values $\chi\cdot\chi$
or $\eta\cdot\eta$
very much larger than $(E_\text{ew})^2$, which is
the case relevant to temperatures far above the phase transition.
Fourth, with a single ultraviolet parameter
$K_{0} \equiv q_{0}/2$, the finite-temperature behavior
\eqref{eq:K-argument-SSB} for the term \eqref{eq:K-argument-KRterm}
essentially gives the previous \textit{Ansatz} \eqref{eq:KAnsatz}.
Again, the phase-transition explanation of the
postulated behavior  \eqref{eq:KAnsatz} does not need to be
taken literally: the origin of $K_T(q)$ may very well
have an entirely different origin,
assuming the model to be relevant at all.

It is clear that
Eqs.~\eqref{eq:Seff-q2def-Fdef}--\eqref{eq:Lscalar-M-rhoV-KAnsatz},
as they stand, only provide a phenomenological model. The main
ingredient is the discontinuous phase-transition-type
behavior of \eqref{eq:KAnsatz}. The model is simpler than the one
used in Ref.~\cite{K2010-Lambda-TeV} and, more importantly,
entirely within the framework of $q$--theory
(which provides a possible solution of CCP1).

%%\newpage%%tmp
\section{Cosmology}
\label{sec:Cosmology}

\subsection{Setup}
\label{subsec:Setup}

A spatially flat, homogeneous, and isotropic universe with scale
factor $a(t)$ will be considered~\cite{Mukhanov2005,PDG2010}.
For convenience, this cosmology will be called a
Friedmann--Robertson--Walker (FRW) universe, even though,
as will become clear in Sec.~\ref{subsec:Numerical-solution},
the Friedmann equation is slightly modified when the
electroweak kick sets in.

The homogeneous matter content of this model universe consists of
two perfect fluids (called type 1 and type 2), with energy density
$\rho_{M1}(t)$ from massive $\sigma$ scalars with an effective
number of degrees of freedom $N_{1}$ and
energy density $\rho_{M2}(t)$ from massless $\psi$ scalars
with $N_{2}$ degrees of freedom
(see the first paragraph of Sec.~\ref{subsec:General-properties}
and the Appendix of Ref.~\cite{K2010-Lambda-TeV}
for the physical motivation of having $N_{1}=N_{2}=10^{2}$).
In thermal equilibrium and without energy exchange,
the type--2 energy density
is given by $\rho_{M2}= (N_{2}\,\pi/30)\,T^4$.
The temperature of the Universe can, therefore, be identified
as approximately $(\rho_{M2})^{1/4}$.

For $\rho_{M1}=0$ and $\rho_{M2} \sim (E_\text{ew})^4$,
the expansion of the Universe
is governed by a Friedmann-type equation (see below)
with a timescale set by
\beq\label{eq:t_ew-def}
t_\text{ew}  \equiv  E_\text{Planck}/(E_\text{ew})^2 \,,
\eeq
in terms of the reduced Planck energy from \eqref{eq:q0def-EPlanck}.
A value $E_\text{ew} \sim \text{TeV}$ gives
$1/t_\text{ew} \sim \text{meV}$.

%%\newpage%%tmp
\subsection{Frozen-electroweak-kick mechanism}
\label{subsec:Frozen-electroweak-kick-mechanism}

With the field-theoretic model of
Sec.~\ref{sec:Field-theoretic-model} for an energy scale
$E_\text{ew} \sim \text{TeV}$,
the basic steps of the frozen-electroweak-kick mechanism
in a flat FRW universe are as follows:
\begin{enumerate}
\item[(i)]
start from a standard radiation-dominated FRW universe
at an ultrahigh temperature $T$ with $\rho_{V}=\rho_{M1}=0$ and
$\rho_{M2} \sim T^4$ from the massless scalars $\psi$
(the $\psi$ scalars have standard
electroweak interactions and are in thermal equilibrium,
whereas the $\sigma$ scalars may have nonstandard interactions
and are assumed to be initially absent, $\rho_{M1}=0$);
\item[(ii)]
as the temperature $T$ drops below $T_{c,\,g}$
(with $T_{c,\,g} \gtrsim E_\text{ew}$), the $\psi^2\,\sigma^2$ coupling
of the scalar theory \eqref{eq:Lscalar} generates a nonzero density
of massive $\sigma$ scalars at cosmic times $t$
around $t_{c,\,g}$ (with $t_{c,\,g} \lesssim t_\text{ew}$);
\item[(iii)]
the presence of massive scalars $\sigma$ modifies the Hubble expansion
rate $H(t)\equiv a(t)^{-1}\,da(t)/dt$ at $t \sim t_\text{ew}$;
\item[(iv)]
the modified Hubble expansion rate kicks
$\rho_{V}$ away from zero~\cite{KV2009-electroweak}, with
$\rho_{V}(t) \sim H(t)^4 \sim (1/t_\text{ew})^4 \sim (\text{meV})^4$
at $t \sim t_\text{ew}$;
\item[(v)]
a nonzero value of $\rho_{V}$ remains when
$K_{T}$ from \eqref{eq:KAnsatz}
is frozen to the constant value $q_0/2$ at a temperature
$T=T_{c,\,K}$ (with $T_{c,\,K} \lesssim E_\text{ew}$)
or cosmic time $t=t_{c,\,K}$ (with $t_{c,\,K} \gtrsim t_\text{ew}$);
\item[(vi)]
the subsequent evolution is that of
a standard $\Lambda$--FRW universe (here, with relativistic
scalars $\psi$, as the massive scalars $\sigma$ ultimately
disappear).\footnote{\label{ftn:QCDcond}Strictly
speaking, such a standard $\Lambda$--FRW
universe would not allow
for further large contributions to the vacuum energy density
at temperatures $T\leq T_{c,\,K} \sim E_\text{ew}$.
Naively, one expects a contribution
of order $(100\;\text{MeV})^4$ from quantum chromodynamics at
$T \sim 10^2\;\text{MeV}$, but it has also been argued that
this is not the case~\cite{BrodskyShrock2009}.}
\end{enumerate}

The phenomenological model of Sec.~\ref{sec:Field-theoretic-model}
is, most likely, over-simplified, but may provide a benchmark
calculation for a dynamically generated cosmological constant.
Expanding on items (iv)--(vi) above,
note that, at cosmic times $t \sim t_\text{ew}$,
the frozen vacuum energy density
$\rho_{V,\,\text{remnant}}\sim (1/t_\text{ew})^{4}$
is negligible compared to the matter energy density
$\rho_{M2}(t_\text{ew})$ by a factor $\xi\sim 10^{60}$.
With $\rho_{M2}(t)\propto 1/t^2$, the tiny (but constant)
vacuum energy density $\rho_{V,\,\text{remnant}}$
only becomes dominant at very much later times,
$t_{VM-\text{equal}} \sim \sqrt{\xi}\:t_\text{ew} \sim 10^{18}\;\text{s}$,
suggesting a possible solution of the so-called cosmic coincidence puzzle
(see, e.g.,
Ref.~\cite{ArkaniHamed-etal2000,Zlatev-etal1998,SahniStarobinsky2000}).

In the next subsection, a preliminary numerical calculation is presented,
which supports the scenario of the frozen-electroweak-kick mechanism
outlined above. As this next subsection is rather technical, it
may be skipped in a first reading, except for a quick look at
the numerical results in Fig.~\ref{fig:1}.

%%\newpage%%tmp
\subsection{Numerical solution}
\label{subsec:Numerical-solution}

With the timescale  $t_\text{ew}$ from \eqref{eq:t_ew-def}
and the hierarchy parameter $\xi$ from \eqref{eq:xi-def}, it turns out
to be useful to introduce the following dimensionless variables
for the cosmic time, the Hubble expansion rate, the energy densities,
and the $q$ shift away from equilibrium~\cite{K2010-Lambda-TeV}:
\bsubeqs\label{eq:dimensionless-var}
\beqa
%\hspace*{-9mm}
\tau &\equiv& (t_\text{ew})^{-1}\, t \,,
\qquad\qquad\;\;\;
h   \equiv  t_\text{ew}\,H \,,
\label{eq:dimensionless-var-tau+h}\\[2mm]
%\hspace*{-9mm}
r_{Mn} &\equiv& \xi^{-1}\, (t_\text{ew})^4\, \rho_{Mn}\,,
\qquad
r_{V} \equiv (t_\text{ew})^4\, \rho_{V} =x^{2}/2 \,,
\label{eq:dimensionless-var-rV+rM}\\[2mm]
%\hspace*{-9mm}
x &\equiv&  \xi\, \big( q/q_0 - 1\big) \,,
\label{eq:dimensionless-var-x}
\eeqa
\esubeqs
where $n$ stands for the matter-species label ($n=1,\, 2$), and the
$\rho_{V}$ \textit{Ansatz} \eqref{eq:rhoV-Ansatz} has been used.
 From now on, an overdot will denote differentiation
with respect to $\tau$.

The relevant dimensionless ordinary differential equations (ODEs)
for the model of Sec.~\ref{sec:Field-theoretic-model}
are then given by~\cite{KV2008-dynamics,K2010-Lambda-TeV}
\bsubeqs\label{eq:ODEs-dimensionless}
\beqa
\hspace*{-0mm}&&
\big(\dot{h} +2h^2 \big)\,
\Big[x^{2}/2 + \xi\,\big(r_{M1} + r_{M2} - 3\, h^2\big)
- 3\, h^2\,x\,\widehat{\theta}\, \Big]
-h\,x\,\dot{x}=0,
\label{eq:ODEs-dimensionless-hdot}
\\[2mm]
\hspace*{-0mm}&&
\dot{r}_{M1}+(4-\overline{\kappa}_{M1})\,h\,r_{M1}
=
\lambdaTwoOne\,r_{M2}-\lambdaOneTwo\,  r_{M1},
\label{eq:ODEs-dimensionless-rM1dot}
\\[2mm]
\hspace*{-0mm}&&
\dot{r}_{M2}+4\,h\,r_{M2}
=
-\lambdaTwoOne\,r_{M2}+\lambdaOneTwo\,r_{M1},
\label{eq:ODEs-dimensionless-rM2dot}
\\[2mm]
\hspace*{-0mm}&&
\big(3\,h\,\dot{x}+ 3\, h^2\,x \big)\;\widehat{\theta}
-\Big[x^{2}/2 + \xi\,\big(r_{M1} + r_{M2} - 3\, h^2\big)\Big]=0,
\label{eq:ODEs-dimensionless-xdot}
\eeqa
\esubeqs
with the EOS function $\overline{\kappa}_{M1}(\tau)$
defined in Sec.~A2 of Ref.~\cite{K2010-Lambda-TeV}
and the effective step function
\bsubeqs\label{eq:widehat-theta}
\beqa
\hspace*{-0mm}&&
\widehat{\theta}(\tau) \equiv \theta\big[r_{M2}(\tau)-r_{c,\,K}\big]\,,
\eeqa
using definition \eqref{eq:theta-def}.
Here, the energy density $r_{M2}$ of the massless scalars $\psi$
monitors the ambient temperature  of the model universe
and determines the moment when the $q$--dependence of $K(q)$
changes from $q/2$ to $q_0/2$, as given by \textit{Ansatz}
\eqref{eq:KAnsatz}. For the benefit of the reader,
the four ODEs in \eqref{eq:ODEs-dimensionless} trace back to
Eqs.~(4.1) and (4.2a) of Ref.~\cite{KV2008-dynamics}
evaluated for the $K_{T}$--\textit{Ansatz}~\eqref{eq:KAnsatz}
and Eqs.~(A5b) and (A5d) of Ref.~\cite{K2010-Lambda-TeV}
adapted to the case considered.

As discussed in Sec.~\ref{subsec:Specific-properties},
the scalar interactions turn on below a
certain critical temperature, which corresponds to
$r_{M2}\leq r_{c,\,g}$ in the cosmological context.
The postulated behavior \eqref{eq:gT-Ansatz} suggests
the following coupling parameters
in the ODEs~\eqref{eq:ODEs-dimensionless}:
\beqa
\lambda_{12}(\tau)&=&\lambda\;\theta[r_{c,\,g}-r_{M2}]\;
\left(\;1-\sqrt{r_{M2}/r_{c,\,g}}\;\right)^2\,,
\label{eq:lambda12}
\\[2mm]
\lambda_{21}(\tau)&=& \lambda_{12}(\tau)
%\nonumber\\&&\times  %%twocolumn
\exp\left[-
\left(\frac{\pi N_{2}}{30\, r_{M2}(\tau_\text{min})}\right)^{1/4} \,
\frac{a(\tau)}{a(\tau_\text{min})}\;\frac{M}{E_\text{ew}}\right]\,,
%\nonumber\\&&  %%twocolumn
\label{eq:lambda21}
\eeqa
\esubeqs
with $\lambda \propto (g_0)^2$.
The argument of the exponential in \eqref{eq:lambda21}
equals the negative inverse of
the $T/M$ expression (A3d) from Ref.~\cite{K2010-Lambda-TeV},
and $\tau_\text{min}$ is an arbitrary reference time
before $\tau_\text{bcs}$ to be introduced below.
The exponential factor of \eqref{eq:lambda21} ensures that
the ODEs \eqref{eq:ODEs-dimensionless-rM1dot}
and \eqref{eq:ODEs-dimensionless-rM2dot} for
Minkowski spacetime ($H=0$) at a fixed temperature $T < T_{c,\,g}$
($r_{M2}< r_{c,\,g}$) give an
equilibrium ratio $r_{M1}/r_{M2} = \exp[-M/T]$.

The new physics from Eqs.~\eqref{eq:gT-Ansatz} and \eqref{eq:KAnsatz}
is assumed to operate in a temperature window set by $r_{M2}(\tau)$
values between $r_{c,\,g}$ and $r_{c,\,K}$, with
\beqa
r_{c,\,g} > r_{c,\,K} \,,
\eeqa
according to \eqref{eq:Tc-assumptions}.
In this regime, the square bracket
in \eqref{eq:ODEs-dimensionless-xdot} corresponds to
the standard Friedmann
equation ($H^2 \propto \rho_\text{tot}$),
to which are added two terms
(proportional to $\dot{q}$ and $q-q_0$)
tracing back to the $q$--dependence of the gravitational-coupling
parameter~\cite{KV2008-dynamics}.

The ODEs \eqref{eq:ODEs-dimensionless}
have two interesting analytic solutions.
The first corresponds to a standard
radiation-dominated FRW universe at high enough temperatures
[with $\lambda_{12}(\tau)=\lambda_{21}(\tau)=0$],
\bsubeqs\label{eq:standard-solutions-FRW-deS}
\beqa
h(\tau) &=& (1/2)\,\tau^{-1}\,,\quad
x(\tau)=r_{M1}(\tau)=0\,,\quad
r_{M2}(\tau)=(3/4)\,\tau^{-2}> r_{c,\,g}\,, %%preprint
%\nonumber\\                                   %%twocolumn
%r_{M2}(\tau)&=&(3/4)\,\tau^{-2}> r_{c,\,g}\,, %%twocolumn
\label{eq:standard-solutions-FRW}
\eeqa
and the second to a standard de-Sitter universe with
constant vacuum energy density and without matter,
\beq
h^2=\xi^{-1}\,x^2/6\,,\quad
\dot{h}=\dot{x}=0\,,\quad
r_{M1}=r_{M2}=0\,.
\label{eq:standard-solutions-deS}
\eeq
\esubeqs
The numerical solution to be presented shortly will be seen to
interpolate between these two analytic solutions.

The hypothetical $\text{TeV}$--scale physics has a very large
hierarchy parameter $\xi\sim 10^{60}$  from \eqref{eq:xi-def},
and four $\xi=\infty$ equations turn out to be relevant
for the phase of the electroweak kick~\cite{K2010-Lambda-TeV}.
Specifically, there are two differential equations,
\bsubeqs\label{eq:ODEs-dimensionless-xi-infty}
\beqa
%\hspace*{-6mm}
&&
\big(\,\widehat{\theta}\;\big)\,h\,\dot{x}\,
\left[3\,  \big(\dot{h} +2h^2 \big)
     - \frac{1}{2}\,\overline{\kappa}_{M1}\,(3\,h^2-r_{M2}) \right]
%\nonumber\\\hspace*{-6mm}&&  %%twocolumn
=
\big( 1-\widehat{\theta}\big)\,h\,x\,\Big\{ \dot{x} \Big\}\,,
\label{eq:ODEs-dimensionless-hdot-xi-infty}
\\[2mm]
%\hspace*{-6mm}
&&
\dot{r}_{M2}+4\,h\,r_{M2}+\lambdaTwoOne\,r_{M2}
=\lambdaOneTwo\,(3\,h^2-r_{M2})\,,
\label{eq:ODEs-dimensionless-rM2dot-xi-infty}
\eeqa
\esubeqs
and two algebraic equations,
\bsubeqs\label{eq:Eqs-dimensionless-xi-infty}
\beqa
%\hspace*{-8mm}
x &=&
\big(\,\widehat{\theta}\;\big)\,
\left[  \frac{1}{2}\,\overline{\kappa}_{M1}\,(3\,h^2-r_{M2}) \right]
%\nonumber\\\hspace*{-8mm}&&  %%twocolumn
+\big( 1-\widehat{\theta}\big)
\left\{ \frac{1}{2} \overline{\kappa}_{M1} (3\,h^2-r_{M2})
\right\}_{r_{M2}=r_{c,\,K}},
\label{eq:Eqs-dimensionless-x-xi-infty}\\[2mm]
%\hspace*{-8mm}
r_{M1} &=&  3\,h^2-r_{M2}\,,
\label{eq:Eqs-dimensionless-rM1-xi-infty}
\eeqa
\esubeqs
with $\widehat{\theta}$ defined by \eqref{eq:widehat-theta}.
Recall that $\overline{\kappa}_{M1}$ is a functional of
$h(\tau) \equiv a(\tau)^{-1}\,da(\tau)/d\tau$,
explicitly given by Eqs.~(A3a)--(A3d) in Ref.~\cite{K2010-Lambda-TeV}.
For the numerical analysis of
Eqs.~\eqref{eq:ODEs-dimensionless-xi-infty}--\eqref{eq:Eqs-dimensionless-xi-infty},
the relevant parts of the equations for
the $\widehat{\theta}=1$ (high-temperature) phase
have been indicated by square brackets, and those for
the $\widehat{\theta}=0$ (low-temperature)
phase by curly brackets, keeping
the two generally valid equations without such brackets.

Returning to general values of $\xi$,
the numerical solution of the ODEs \eqref{eq:ODEs-dimensionless}
has been obtained for the case with $N_{1}=N_{2}=10^2$
and an equal mass $M$ of all type-1 particles
(i.e., the case-B mass spectrum in the terminology
of Sec.~A2 of Ref.~\cite{K2010-Lambda-TeV}).
As mentioned above, at temperatures
above the critical temperature $T_{c,\,g}$, it is possible to have
a standard radiation-dominated FRW universe
\eqref{eq:standard-solutions-FRW} with all type--2
particles in thermal equilibrium
(the type--1 particles are assumed to be absent in the early phase,
see Sec.~\ref{subsec:Frozen-electroweak-kick-mechanism}).
Hence, the appropriate boundary conditions on the four dynamical
variables at a time $\tau=\tau_\text{bcs}$ are
\bsubeqs\label{eq:ODEs-dimensionless-bcs}
\beqa
h(\tau_\text{bcs})&=& \frac{1}{2} \,(\tau_\text{bcs})^{-1} \,,
\\[2mm]
r_{M1}(\tau_\text{bcs})&=& 0\,,
\\[2mm]
r_{M2}(\tau_\text{bcs})&=& 3\;\big[h(\tau_\text{bcs})\big]^2\,,
\\[2mm]
x(\tau_\text{bcs})&=& 0 \,.
\eeqa
\esubeqs
The precise value of $\tau_\text{bcs}$ is
irrelevant as long as it is sufficiently small,
with $r_{M2}(\tau_\text{bcs})\geq r_{c,\,g}$
[physically interpreted as $T(t_\text{bcs})\geq T_{c,\,g}$].
Furthermore, choose for $\lambda$ the value $10^{4}$, making
$r_{M1}(\tau)$ decrease significantly before the $K_{T}$ transition
at $r_{M2}=r_{c,\,K}$
is reached (the new physics then operates in a relatively narrow
temperature interval near $T\sim E_\text{ew}$). Other values
$\lambda\gtrsim 10^{3}$   %%FRK-num-v202 %%FRK-num-v3998
give similar numerical solutions.

%\newpage
\begin{figure*}[t]
%%do not use eps2eps at ITP,
%%rename fig1tmp_v3->fig1_v3->fig1_v4
\hspace*{-6mm}
\includegraphics[width=1.06\textwidth]{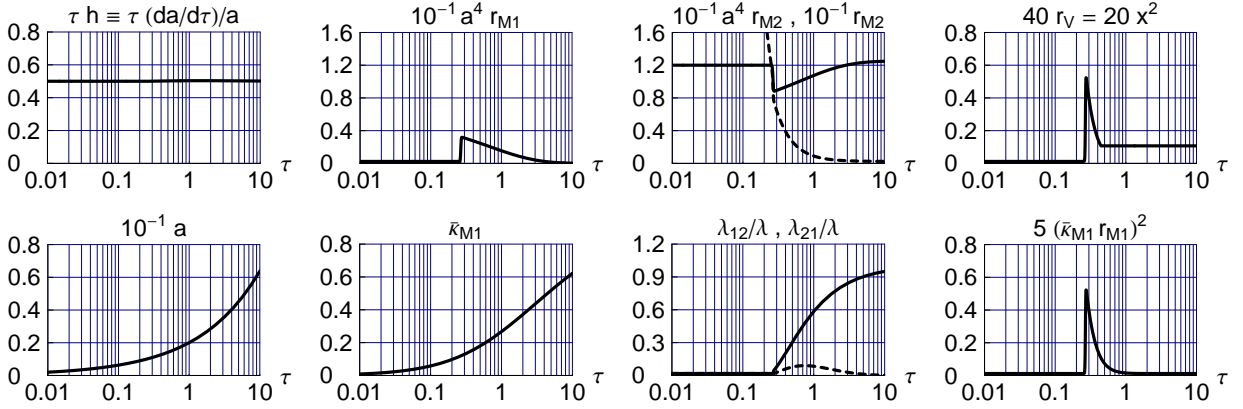}
\vspace*{-2mm}
\caption{Numerical solution of the dimensionless ODEs
\eqref{eq:ODEs-dimensionless}
with EOS function $\overline{\kappa}_{M1}(\tau)$
defined in Sec.~A2 of Ref.~\cite{K2010-Lambda-TeV}
and further parameters \eqref{eq:widehat-theta}.
The panels are organized as follows: the four basic dynamic variables
[$h(\tau)$, $r_{M1}(\tau)$, $r_{M2}(\tau)$, and $x(\tau)$
as defined by \eqref{eq:dimensionless-var}] are shown on the top row
and secondary or derived quantities on the bottom row.
The dashes lines in the panels of the third column
correspond to $10^{-1}\,r_{M2}$ and $\lambda_{21}/\lambda$.
The main result is the nonzero remnant value of the dimensionless
gravitating vacuum energy density $r_{V}\equiv x^{2}/2$
shown in the top right panel. The model parameters are
$\{\xi,\,  \lambda ,\, r_{c,\,g},\, r_{c,\,K} \}$ $=$
$\{10^7,\,  10^4   ,\, 12 ,\,       3\}$.  %%FRK-num-v202  %%FRK-num-v3998
The ODEs are solved over the interval
$[\tau_\text{min},\, \tau_\text{max}]$ $=$ $[0.01,\, 10]$
with the following boundary conditions \eqref{eq:ODEs-dimensionless-bcs}
at $\tau=\tau_\text{bcs}=0.25$ corresponding to
$r_{M2}(\tau_\text{bcs})=r_{c,\,g}\,$:
$\{x,\, h,\,       a,\,  r_{M1},\, r_{M2}\}$ $=$
$\{0,\, 2,\,       1,\,  0,\,      12\}$.}  %%FRK-num-v202 %%FRK-num-v3998
\label{fig:1}
\end{figure*}
%
%\vspace*{3cm}\begin{table*}[h]  %%preprint
\begin{table}  %%twocolumn
%\begin{center} %%preprint
\caption{Asymptotic values of the dimensionless
gravitating vacuum energy density $r_{V}(\tau)$
for various hierarchy parameters $\xi$ ranging from $10^4$ to $10^8$.
All other parameters are given in the caption of Fig.~\ref{fig:1}.
The entry for $\xi=\infty$ has been calculated from
Eqs.~\eqref{eq:ODEs-dimensionless-xi-infty} and
\eqref{eq:Eqs-dimensionless-xi-infty}.
The numerical accuracy is estimated at $\pm 1$ in the
last digit shown.\vspace*{2mm}}
\label{tab-xi-rV}
\renewcommand{\tabcolsep}{3pc}    %% enlarge column spacing
\renewcommand{\arraystretch}{1.1}   %% enlarge line spacing
\begin{tabular}{c|c}
\hline\hline
 $\xi$ & $10^3 \times r_{V}(\infty)$\\
\hline
 $10^{4}$ & $ 7.310 $\\ %%7.3103583478672  %%FRK-num-v3998
 $10^{5}$ & $ 5.380 $\\ %%5.3799088689754
 $10^{6}$ & $ 2.028 $\\ %%2.028324666188488
 $10^{7}$ & $ 2.376 $\\ %%2.37622989607275
 $10^{8}$ & $ 2.376 $\\ %%2.37639796278357
 $\infty$ & $ 2.376 $\\ %%  unchanged
\hline\hline
\end{tabular}
%\end{center} %%preprint
%\end{table*} %%preprint
\end{table}  %%twocolumn

%\newpage %%preprint
The numerical results are shown in Fig.~\ref{fig:1}. The $r_{V}$ panel,
in particular, shows the narrow new-physics window
with the critical temperature \eqref{eq:gT-Ansatz} at $\tau = 0.25$
(from $r_{M2}=r_{c,\,g}$) and the freezing of the gravitational-coupling
parameter \eqref{eq:KAnsatz} at
$\tau \sim 0.45$ (from $r_{M2}=r_{c,\,K}$).  %%FRK-num-v202 %%FRK-num-v3998
The calculated numerical value of $r_{V,\text{\,remnant}}$ is approximately
$2.4 \times 10^{-3}$.   %%FRK-num-v202  %%FRK-num-v3998
For different values of $r_{c,\,K}$
than chosen in  Fig.~\ref{fig:1}, while keeping the other
parameters the same, the numerical values of $r_{V,\text{\,remnant}}$
will, of course, be less than the maximal value of $r_{V}(\tau)$
shown in the top right panel of the figure, that is,
$r_{V,\text{\,remnant}}\lesssim 1.5 \times 10^{-2}$.  %%FRK-num-v202 %%FRK-num-v3998
Incidentally, the peak of the $r_{V}(\tau)$ curve at
$\tau = \tau_\text{peak} \sim 0.275$ %%FRK-num-v202 %%FRK-num-v3998
looks sharp in the plot but is really
a concave parabola with a large negative second derivative,
$r_{V}''(\tau_\text{peak}) \sim -10^2$.  %%FRK-num-v202 %%FRK-num-v3998

Table~\ref{tab-xi-rV} presents the numerical values for
$r_{V}(\infty)\equiv \lim_{\tau\to\infty} r_{V}(\tau)$
for various hierarchy parameters $\xi$. The $r_{V}(\tau)$ solution
for $\xi\lesssim 10^6$ has, %%FRK-num-v202  %%FRK-num-v3998
in fact, significant oscillations superposed on the smooth
curve shown in the bottom right panel of Fig.~\ref{fig:1},
which explains the somewhat erratic behavior of the first three
entries in Table~\ref{tab-xi-rV}. Based on the analysis of
Ref.~\cite{K2010-Lambda-TeV}, the results for $\xi=\infty$ can be
expected to give a close approximation to those for $\xi\sim 10^{60}$,
which corresponds to the physically relevant case
according to \eqref{eq:xi-def}.

%%\newpage%%tmp
\subsection{Analytic result for $\boldsymbol{\xi=\infty}$}
\label{subsec:Analytic result}

The $\xi=\infty$ equations~\eqref{eq:ODEs-dimensionless-xi-infty}
and \eqref{eq:Eqs-dimensionless-xi-infty} immediately
give an analytic expression for the asymptotic value
of the dimensionless vacuum energy density,
\beqa\label{eq:rVfreeze-analytic}
%\hspace*{-11mm}&&
\lim_{\tau\to\infty} r_{V}(\tau)\,\Big|^{\xi=\infty}
=
\frac{1}{8}\,
\Big(
\overline{\kappa}_{M1}(\tau_\text{freeze})\,
%\nonumber\\\hspace*{-11mm}&&\times %%twocolumn
\big[3\,h(\tau_\text{freeze})^2-r_{M2}(\tau_\text{freeze})\big]
\Big)^2\,
\Big|_{\,r_{M2}(\tau_\text{freeze})=r_{c,\,K}}\,,
\eeqa
where the EOS function $\overline{\kappa}_{M1}(\tau)$ is determined by
the dimensionless Hubble expansion rate $h(\tau)$
and $r_{Mn}(\tau)$ stands
for the dimensionless energy density of matter component $n$
[all dimensionless variables are defined in
\eqref{eq:dimensionless-var}].

Expression \eqref{eq:rVfreeze-analytic} is formal,
because the numerical solution of the two
nonlinear ODEs \eqref{eq:ODEs-dimensionless-hdot-xi-infty} and
\eqref{eq:ODEs-dimensionless-rM2dot-xi-infty} for $\widehat{\theta}=1$
is needed to determine the value of, for example,
$h(\tau_\text{freeze})$.
But the analytic expression \eqref{eq:rVfreeze-analytic}
does clarify the essential physics involved:
the EOS function $\overline{\kappa}_{M1}(t)$
multiplied by the corresponding energy density of massive particles
($r_{M1}= 3\,h^2-r_{M2}$)
and the freezing of $K(t)=q(t)/2$ to
the constant value $q_0/2$ at a cosmic temperature of the
order of $E_\text{ew} \sim \text{TeV}$.

%%\newpage%%tmp
\section{Conclusion}
\label{Conclusion}

In this article, a simple field-theoretic model
\eqref{eq:Seff-q2def-Fdef}--\eqref{eq:Lscalar-M-rhoV-KAnsatz}
has been presented, which remains entirely within the framework of
$q$--theory~\cite{KV2008-statics,KV2008-dynamics,KV2009-CCP1}
and does not require unnaturally small coupling
constants (the energy scales $E_\text{ew}\sim \text{TeV}$
and $E_\text{Planck}\sim 10^{15}\;\text{TeV}$
are considered to be given~\cite{ArkaniHamed-etal2000}).
As summarized in Sec.~\ref{subsec:Frozen-electroweak-kick-mechanism},
the model generates via the
electroweak-kick mechanism~\cite{KV2009-electroweak}
an effective cosmological constant $\Lambda_\text{eff}$
(remnant vacuum energy density $\rho_{V}$), which is
consistent with the value
$\Lambda^\text{obs} \sim 1\times 10^{-29}\;\text{g}\,\text{cm}^{-3}
\sim (2\times 10^{-3}\;\text{eV})^4$
from observational cosmology~\cite{SahniStarobinsky2000,PDG2010}.
In addition, having $\Lambda_\text{eff}\sim
(E_\text{ew})^8/(E_\text{Planck})^4$ provides a natural
explanation~\cite{ArkaniHamed-etal2000}
of the fact that the orders of magnitude of the energy densities of
vacuum, matter, and radiation are approximately the same in the
present Universe (also known as the triple cosmic coincidence puzzle).

With the calculated dimensionless vacuum
energy density $r_{V}(\infty) \sim 2.4 \times 10^{-3}$  %%FRK-num-v202 %%FRK-num-v3998
from Table~\ref{tab-xi-rV},
the required energy scale $E_\text{ew}$ of the new physics
is approximately $4.7\;\text{TeV}$,                    %%FRK-num-v202 %%FRK-num-v3998
according to Eq.~(5.2) of Ref.~\cite{K2010-Lambda-TeV}.
However, the main focus of the present article is not on numerical
estimates (plenty have been given in Ref.~\cite{K2010-Lambda-TeV}),
but rather on the physical content of a
theory capable of generating the observed cosmological
``constant'' of our Universe.

In that spirit, the most interesting result of this article
is the observation that the proposed model involves one
crucial ingredient, namely, the discontinuous
phase-transition-type behavior of the dependence
of the gravitational coupling \eqref{eq:KAnsatz}
on the quinta-essentia field $q$.
(The need for some form of singular behavior in order to freeze
$\rho_{V}[q(t)]$ has also been emphasized in the second remark of
Sec.~III E of Ref.~\cite{K2010-Lambda-TeV}.) The main task is,
therefore, to find the rationale for this phase-transition type of
behavior or for a different effect with the same result of
freezing part of the vacuum energy density
below a certain cosmic temperature scale.
Furthermore, the simple model (or a suitable generalization of it)
needs to be embedded in the complete solution of the
cosmological constant problem, which is still outstanding.

\section*{\hspace*{-4.5mm}ACKNOWLEDGMENTS}
\vspace*{-0mm}\noindent
The author gratefully acknowledges the hospitality of
the Perimeter Institute for Theoretical Physics, Canada,
where this work was initiated.
He also thanks M. Guenther, G.E. Volovik, and the referee
for helpful comments on an earlier version of this article.

%%\newpage%%tmp

\end{document}